\documentclass{aa}

\usepackage{txfonts}
\usepackage{graphicx}
\usepackage{lscape}
\usepackage{longtable}
\usepackage{natbib}

\begin{document}

\title{Accurate fundamental parameters and distance to \\a massive early-type eclipsing binary \\ in the Danks 2 cluster\thanks{Based on observations made with ESO Telescopes at the La Silla Paranal Observatory under program ID 090.D-0065(A).}}
\author{M. Kourniotis\inst{1,2,}\thanks{mkourniotis@astro.noa.gr} \and ~~A.Z. Bonanos\inst{1} \and S.J. Williams\inst{1} \and N. Castro\inst{3} \and E. Koumpia\inst{4} \and J.L. Prieto\inst{5,6}}

\institute{IAASARS, National Observatory of Athens, GR-15236 Penteli, Greece \and Section of Astrophysics, Astronomy and Mechanics, Faculty of Physics, University of Athens, Panepistimiopolis, GR15784 Zografos, Athens, Greece \and Argelander-Institut für Astronomie der Universität Bonn, Auf dem Hügel 71, 53121, Bonn, Germany \and SRON Netherlands Institute for Space Research, Landleven 12, 9747 AD Groningen, The Netherlands; Kapteyn Institute, University of Groningen, The Netherlands \and N\'{u}cleo de Astronom\'{i}a de la Facultad de Ingenier\'{i}a, Universidad Diego Portales, Av. Ej\'{e}rcito 441, Santiago, Chile \and Millennium Institute of Astrophysics, Santiago, Chile} 

\date{}
\authorrunning{Kourniotis et al.}
\titlerunning{Precise distance to the Danks 2 cluster}

\abstract
{We present a study of the properties of the O-type, massive eclipsing binary 2MASS J13130841-6239275 located in the outskirts of the Danks 2 cluster in the G305 star-forming complex, using near-infrared spectroscopy from VLT/ISAAC. We derive the masses and radii to be $24.5\pm0.9$ M$_{\sun}$ and $9.2\pm0.1$ R$_{\sun}$ for the primary and $21.7\pm0.8$ M$_{\sun}$ and $8.7\pm0.1$ R$_{\sun}$ for the secondary component. In addition, we evaluate the sensitivity of our parameters to the choice of the spectral features used to determine the radial velocities. Both components appear to be main-sequence O6.5$-$O7 type stars at an age of $\sim5$ Myr, which is in agreement with the age of the cluster. A high visual extinction of $A_{5495}=11.9\pm0.1$ mag is reported, which is likely attributed to the cold molecular gas contaminating the north-east region of the cluster. By fitting the spectral energy distribution of the system to the available $BVI_{c}JHK_{s}$ photometry, we determine a distance to the system of $3.52\pm0.08$ kpc with a precision of $2\%$, which is the most well-determined distance to the Danks 2 cluster and the host complex reported in the literature.}


\maketitle 

\section{Introduction}
\label{section:intro}

The G305.4+0.1 complex (hereafter, G305) is one of the most massive, ongoing star-forming regions in the Galaxy \citep{Clark04}. It is home to nine \ion{H}{II} regions \citep{Caswell87}, which are powered by the ionizing flux of hot stars and are spatially associated with maser emission and at least five star clusters \citep{Dutra03} embedded in the complex. Whereas the majority of these clusters are compact and thus unresolved, the Danks 1 \& 2 clusters \citep{Danks83, Danks84}, which reside in the center of the complex, are resolved and thus allow for studies of their physical properties. Both clusters have been shown to be $\le7$ Myr old, with Danks 2 being noticeably older than Danks 1 \citep{Chene12}. In addition, initial masses of 8\,000 and 3\,000  M$_{\sun}$ were derived for Danks 1 and 2 respectively, with an uncertainty that reaches $\sim30\%$ \citep{Davies11}. Constraining the properties of the clusters relies on an accurate determination of the distance, extinction and their associated uncertainties. To date, the most robust method to provide accurate distance measurements to young galactic clusters and nearby galaxies is with the use of eclipsing binaries (EBs) \citep[e.g.][]{Pietr13}. 

EBs exhibiting double-lined spectra provide a geometrical method to precisely measure the fundamental parameters of their components \citep{Andersen91,Torres10}. In particular, the light curve provides the orbital period, eccentricity, the flux ratio of the two stars, fractional radii and the inclination of the system with respect to the observer. From the double-lined spectra, we measure the velocities of the individual components. The fit to these velocities yields the velocity semi-amplitudes, and therefore, the ratio of the masses. Effective temperatures and thus luminosities can be estimated by fitting synthetic spectra to the observed ones. Therefore, an accurate determination of the reddening law and thereby distance, is feasible \citep[e.g.][]{Bonanos06, Bonanos11}. Selecting systems that host well-separated components (i.e. detached systems), avoids complexities that originate from a common-envelope state. Such systems are of major importance for testing theoretical stellar evolutionary models of single stars. Light curves of detached systems that display eclipses of similar depth ensure roughly equal contribution of both components to the total flux and such systems with orbital periods $\le10$ days are likely to provide double-lined and well-separated spectra.  

The majority of hot, massive stars are claimed to be in close binary systems \citep{Chini12,Sana12,Sota14}, thus increasing the possibility of observing short-period, eclipsing configurations \citep[e.g.][]{Kourniotis14}. Such systems not only constitute extragalactic anchors, ideal for the calibration of the cosmic distance scale \citep[e.g.][]{Guin98,Hild05,Ribas05,Bonanos06,Bonanos11}, but also, when found in young, Galactic clusters, can set tight constrains on the age, chemical composition and distance \citep[e.g.][]{South06,Koumpia12}. In this work, we investigate the properties of 2MASS J13130841-6239275 (hereafter, D2-EB), a massive EB located $\sim$$2\arcmin$ from the center of the Danks 2 cluster that constitutes a luminous, double-lined spectroscopic system. A summary of the parameters of D2-EB derived in the present work along with the parameters of the Danks 2 cluster, are listed in Table \ref{tab01}.

D2-EB\footnote{RA = 13:13:08.41, Dec = $-$62:39:27.5, J2000} was discovered to be a well-detached eclipsing system during the optical variability survey by Bonanos et al. (in prep.). Due to the heavy foreground visual extinction in the direction of the clusters \citep[A$_{v}=8-9$ mag,][]{Bica04}, we obtained near-infrared \textit{K}-band spectroscopy to resolve double lines in the system. The paper is organized as follows: in Sect. \ref{obs} we describe the follow-up spectroscopy and the data reduction, in Sect. \ref{rv} we present the radial velocity analysis, in Sect. \ref{phoebe} we model the binary, in Sect. \ref{dist} we derive the distance, in Sect. \ref{disc} we discuss our results and in Sect. \ref{concl} we finish with some concluding remarks.

\section{Observations and data reduction}
\label{obs}

Bonanos et al. (in prep.) undertook an optical variability survey of the Danks clusters in search of massive EBs over 25 nights in March, April and May 2011 with the 1-m Swope telescope at Las Campanas Observatory in Chile. The telescope is equipped with a $2048\times3150$ detector with a pixel scale of $0.435\arcsec$/pixel and a field of view of $14.8\arcmin\times22.8\arcmin$. Observations were taken between HJD 2\,455\,648.5 and 2\,455\,687.83, in the broadband $B,V,R,I_{c}$ filters over approximately 50, 500, 950 and 530 epochs, respectively. The data were reduced, calibrated to the standard system and the image subtraction package ISIS \citep{Alard98,Alard00} was used to extract the $I_{c}$-band light curves of all sources. The light curves were searched for periodicity with the Analysis of Variance (AoV) algorithm \citep{Schwar89} yielding 21 EBs. Of these, 13 were found to display red colors of $R-I_{c}\ge0$, likely corresponding to members of the highly extincted clusters. The $I_{c}$-band light curve of D2-EB was found to display a strong periodic signal of $\sim3.37$ days and appeared to be an ideal target for follow up spectroscopy: a luminous, well-detached and nearly equal-depth eclipsing system.

\subsection{Spectroscopy}
\label{spec}
We obtained \textit{K}-band spectroscopy of the D2-EB with the Infrared Spectrometer and Array Camera (ISAAC) mounted at the Nasmyth A focus of the 8.2m UT3 (Melipal) at the VLT facility in Chile. The observations were taken under ESO program ID 090.D-0065(A). The spectrograph is equipped with a 1024 x 1024 Hawaii Rockwell array with a pixel scale of $0.148\arcsec$/pixel \citep{Moor98}. A slit width $0.3\arcsec$ was used, resulting in a resolving power of $R\sim7\,500$ as measured from the full width half maximum ($\sim2.88$ \AA) of an unblended telluric OH emission feature at $\sim2.125$ $\mu$m.  To derive a precise measurement of the velocity semi-amplitudes, we requested eight visits, four per quadrature. In total, eight observing blocks were executed in service mode, six of which correspond to the first and two to the second quadrature of the system. For each observing block, nodding $60\arcsec$ along the slit was applied in an AB pattern, resulting in two spectra of total integration time 100 sec each. 

The spectra were reduced using basic IRAF routines\footnote{IRAF is distributed by the National Optical Astronomy Observatories, which are operated by the Association of Universities for Research in Astronomy, Inc., under cooperative agreement with the National Science Foundation.} and wavelength calibrated using telluric OH sky lines \citep{Rous00} that span the \textit{K}-band wavelength range. Subsequently, the extracted pair of spectra of each observing block were normalized and averaged, yielding a signal-to-noise ratio (S/N) $\sim150$ as measured in the middle of the wavelength coverage ($\sim2.14$ $\mu$m). Cosmic rays were rejected using the L.A.Cosmic script \citep{Dok01}. We employed the newly released \textit{molecfit} tool \citep{Smette15}, which generates a synthetic model of atmospheric transmission based on the information provided by the FITS header of the science object, to remove telluric features. Heliocentric corrections were calculated using IRAF's \textit{rvcorr} task and applied. Table \ref{tab02} lists the Heliocentric Julian Dates (HJD) of the observations, exposure times and the values of airmass. Of the eight spectra, we discarded the one obtained on HJD 2\,456\,378.808\,74 from further analysis, as it was taken near an eclipse, at phase $\sim0.47$, and contained only one set of spectral features. 

\section{Radial velocity analysis}
\label{rv}

The spectrum of the D2-EB clearly displays \ion{He}{ii} (2.189 $\mu$m) in absorption, which is prominent for both components in phases outside of eclipse. As in the optical, \ion{He}{II} lines appear only in hotter stars and are the main indicator for identifying O-type stars. The observed \ion{He}{I} (2.112 $\mu$m) in absorption implies that both stars must be later than O5 \citep{Hanson96}. Both components of the D2-EB are hence classified as intermediate/late O-type stars. In addition, our spectra show Br$\gamma$ (2.166 $\mu$m) in absorption, which provides evidence against a supergiant classification. However, a precise luminosity classification strongly depends on the resolution and requires a S/N > 150. Emission of \ion{C}{IV} ($2.08$ $\mu$m) is evident only for the redshifted component as the blueshifted feature is not within the observed wavelength range. Further analysis of the observed features that are used to measure the radial
velocities and to assign a spectral type is presented in Sec. \ref{hanson}.

\subsection{Analysis Using Model Atmospheres}
\label{fw}

\begin{figure*}
\centering 
\includegraphics[width=5in]{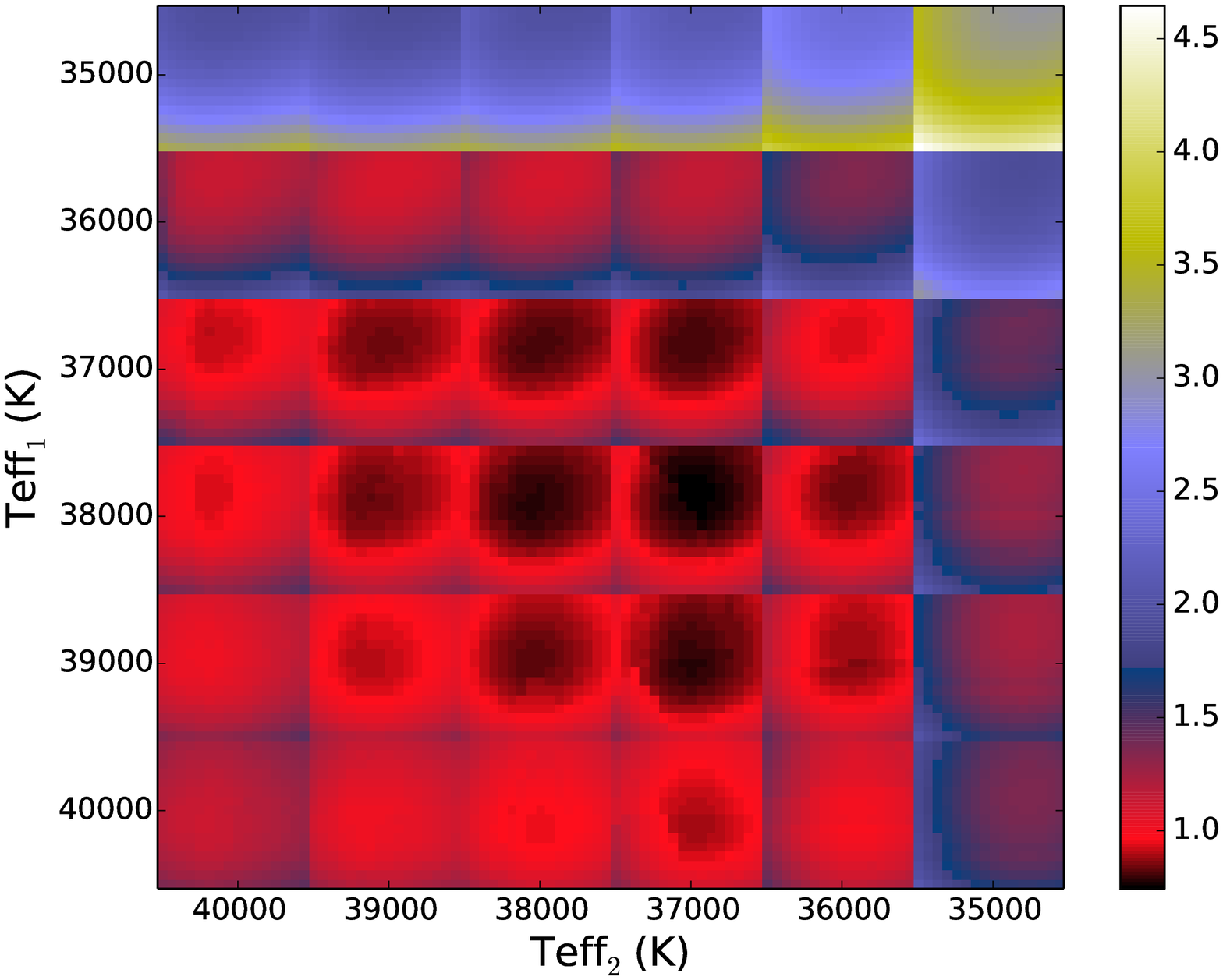}
\caption{Reduced-$\chi^2$ panel of the fit of composite FASTWIND spectra to the ISAAC spectrum of HJD 2\,456\,345.856\,71. Axes denote the effective temperature of the components and are split into six, equal intervals with 16 subdivisions of 10 km~s$^{-1}$ each, to represent the projected rotational velocity range $50-200$ km~s$^{-1}$. We find that the best fit corresponds to a pair of templates with T$_{eff1}=38$ kK, T$_{eff2}=37$ kK and $v\sin i$ values of $150\pm20$ and $130\pm20$ km~s$^{-1}$, respectively.}
\label{fig01}
\end{figure*}

We used the stellar atmosphere code FASTWIND \citep{Santolaya97,Puls05} to synthesize the lines of H and He and measure the radial velocities. The code enables non-LTE calculations and assumes a spherical symmetry geometry. The wind velocity structure is included in the model through a $\beta$-like law. We generated a grid of FASTWIND templates at the metallicity of the Galaxy, spanning the temperature range 35--40 kK with a step of 1 kK and fixed the surface gravity at log $g=3.9$ dex, as inferred from Sect. \ref{phoebe}. The templates were broadened and resampled to match the resolution ($R\sim7\,500$) and dispersion of the observed spectra. To constrain the projected rotational broadening, we implemented a code based on $\chi^2$ minimization setting six free parameters, three for each component; temperature, radial and projected rotational velocity. We built synthetic spectra and fit the regions around the Br$\gamma$ (2.166 $\mu$m)/ \ion{He}{I} (2.162 $\mu$m) blend and \ion{He}{II} (2.189 $\mu$m) feature of the ISAAC spectrum taken on HJD 2\,456\,345.856\,71 that exhibits well-separated spectral features. Each composite spectrum consists of two FASTWIND templates weighted according to the light ratio provided by the light curve analysis and broadened at the projected rotational velocity range $50-200$ km~s$^{-1}$. For the fit, we adopted a $\sigma$ value of 0.006\,7, which corresponds to the standard deviation of the noise in units of normalized flux, based on the mean S/N ratio $\sim150$. Fig. \ref{fig01} displays a panel of our analysis providing the reduced-$\chi^2$ of the fit with respect to the temperatures and projected rotational velocities of the components. For each set of parameters, radial velocities were measured to optimize the fit. We found that the projected rotational velocities that correspond to the best fit are $150\pm20$ km~s$^{-1}$ and $130\pm20$ km~s$^{-1}$ for the primary and the less luminous secondary component, respectively. These values are in good agreement with those of a tidally locked system whose orbital period is identical to the rotational period for both components (Sect. \ref{phoebe}). 

\begin{figure*}
\centering 
\includegraphics[width=5in]{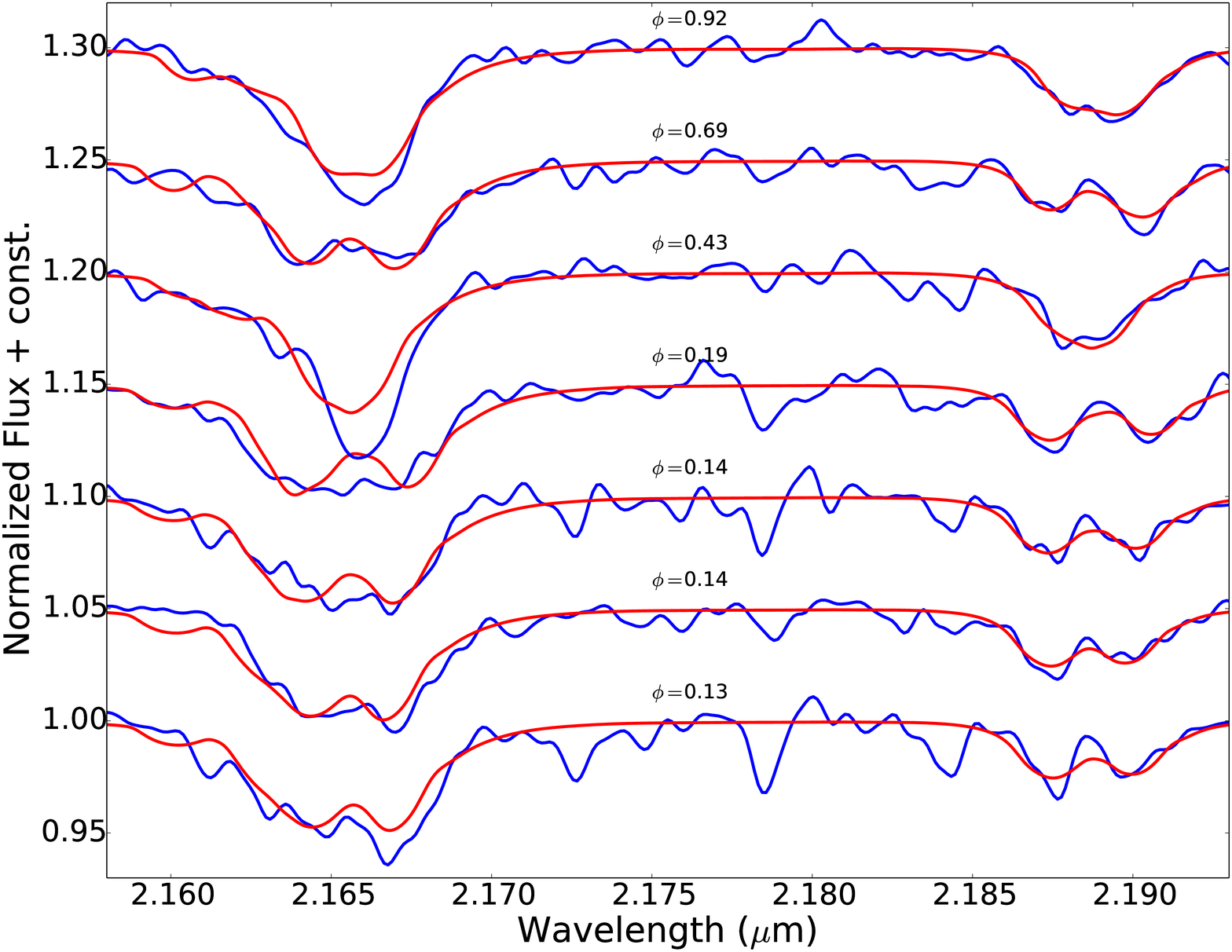}
\caption{Normalized ISAAC spectra (blue) sorted by the orbital phase ($\phi$) of D2-EB, including the Br$\gamma$ (2.166 $\mu$m)/\ion{He}{I} (2.162 $\mu$m) blend and the \ion{He}{II}  (2.189 $\mu$m) line. The composite spectra (red) of two FASTWIND templates of 38kK and 37kK, weighted, summed and shifted to our derived velocities, are overplotted. The best-fit velocites are obtained from the fit of \ion{He}{II}. For clarity, a smoothing window is applied to the observed spectra and a constant is added to the relative fluxes. Absorption features in the wavelength range $2.170-2.185$ $\mu$m are due to poor telluric correction.}
\label{fig02}
\end{figure*}

To construct the radial velocity curve, we fixed the aforementioned projected rotational velocities and ran the fitting algorithm for the seven ISAAC spectra. Given that the best-fit temperatures are found to range from 36 kK to 38 kK with a 1 kK difference between the temperature of the components, we eventually examined two cases: a set with 37 kK and 38 kK models and a set with 36 kK and 37 kK models. While the former resulted in a best fit for the \ion{He}{II} line, the latter appeared to optimally fit the Br$\gamma$/\ion{He}{I} feature. We adopted the first case for providing reliable velocity measurements as \ion{He}{II} is a more accurate tracer of the stellar photosphere for stars at these temperatures. Contours of $\chi^2$ values for the \ion{He}{II} fit were plotted in two dimensional radial velocity diagrams and the center of the contour corresponding to the lowest $\chi^2$ was adopted to generate the best fit. Fig. \ref{fig02} presents the best-fit composite spectra overplotted onto the observed ones, which were smoothed for clarity. The inferred velocities are listed in Table \ref{tab03}. The uncertainties correspond to the standard deviation of the residuals of the observed velocities compared to the best-fit model described in Sect. \ref{phoebe}.

\subsection{Analysis Using Near-Infrared Atlas}
\label{hanson}

We also explored a different method of analysis by comparing our observations against near-IR spectra of early-type stars of similar resolution and S/N. The work by \cite{Hanson05} (hereafter, H05) provides a near-infrared atlas of Galactic hot stars that comprises of 37 well known, OB-type stars observed with VLT/ISAAC and Subaru/IRCS with a resolution of $R\sim8\,000-12\,000$ and a mean S/N $\sim100-300$. The spectral types range from O3 to B3 and sample most luminosity classes. The advantage of using the atlas over the FASTWIND templates lies in the availability of extra diagnostic lines such as \ion{C}{IV} at $2.08$ $\mu$m and the blend of the CNO complex at $2.115$ $\mu$m with \ion{He}{I} at $2.112$ $\mu$m. As mentioned in H05, the atlas should not be used for an accurate spectral classification but rather, to obtain a first estimate for the temperatures. 

Given that the H05 spectra are calibrated to air wavelengths (M. Hanson, private communication), we first applied an air-to-vacuum unit conversion. The resolution of the spectra was reduced to match that of our data and the light ratio was fixed in accordance with the light curve analysis. We found that a composite spectrum of two identical O6.5 III spectra yielded an optimal fit to our observed spectra. However, there is no O6.5 main-sequence counterpart in the atlas, hence luminosity class at that temperature is not well-determined. The resulting spectral type is in agreement with the effective temperature of 38 kK adopted using FASTWIND models; the recent temperature-spectral type calibration for O-type stars by \cite{Simon14} assigns an effective temperature of 38.7 kK to an O6.5 Milky Way dwarf of surface gravity 3.8 dex, with typical uncertainties of up to 1.5 kK and 0.15 dex, respectively. 

\begin{figure*}
\centering 
\includegraphics[width=5in]{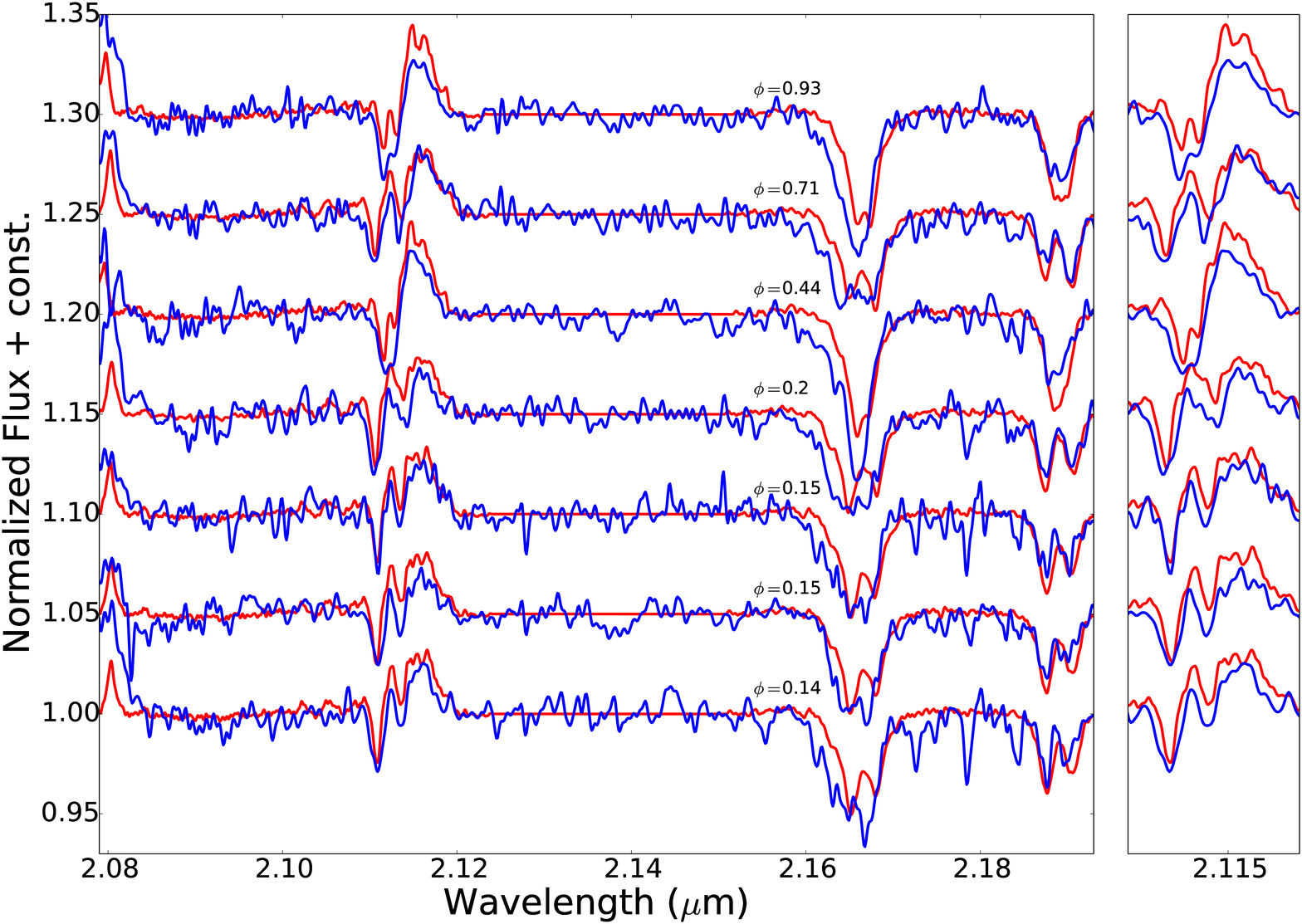}
\caption{Same as in Fig. \ref{fig02}, but the composite spectrum is comprised of two O6.5 III spectra taken from \cite{Hanson05}. In the \textit{left panel}, we show the entire wavelength range of the ISAAC spectra, which includes lines of \ion{C}{IV}  (2.08 $\mu$m), \ion{He}{I} (2.113, 2.162 $\mu$m), the CNO complex (2.115 $\mu$m), Br$\gamma$ (2.166 $\mu$m) and \ion{He}{II}  (2.189 $\mu$m). The best-fit velocities are measured by fitting all features except for the Br$\gamma$/\ion{He}{I} blend. In the \textit{right panel}, we show the result of the separate fit to the CNO/\ion{He}{I} blend, which was employed to derive velocities.}
\label{fig03}
\end{figure*}

The O6.5 III spectrum from H05 displays helium and hydrogen features in both the \textit{H} and \textit{K}-band. While the centroid of the \ion{He}{I} lines appear within 0.5 $\AA$ from their theoretical wavelengths (in air, $1.700\,25$ $\mu$m and $2.112\,01$ $\mu$m), the \ion{He}{II} line ($2.188\,52$ $\mu$m) clearly displays a $\sim1.5$ $\AA$ offset redward. Comparing to the adopted FASTWIND template, the blue wing of \ion{He}{II} appears weaker with respect to the red wing. It is unlikely that this is due to inefficient telluric correction, as the \textit{H}-band \ion{He}{II} feature ($1.691\,84$ $\mu$m) also suffers a similar, although smaller ($\sim1$ $\AA$) offset. In addition, a phase-dependent velocity discrepancy between the CNO/\ion{He}{I} blend and the \ion{He}{II} line is prominent in the ISAAC spectra, as can be seen in Fig. \ref{fig03}. The \ion{He}{II} line is mainly formed in the transition region from the stellar photosphere to the wind \citep{Lenor04}, hence it is dominated by the uncertainty in the wind density structure. We therefore conclude that contamination by stellar winds is the most plausible explanation for this \ion{He}{II} offset (A. Herrero, private communication) as stellar winds are prominent in early-type stars. To avoid a definite bias of this effect to our study, we proceeded to measure velocities with the H05 spectra in two ways; first, fitting all features except for the Br$\gamma$/\ion{He}{I} blend and second, fitting only the $2.112$ $\mu$m \ion{He}{I} line along with the CNO complex. While the first method relies on the contribution of four independent features, it is hindered by the uncertainty of the aforementioned \ion{He}{II} effect and of a possible, inefficient correction in our observations for the strong, telluric CO$_{2}$ absorption band near $2.08$ $\mu$m, in the region of \ion{C}{IV}. Furthermore, both \ion{C}{IV} and \ion{He}{II} are found at the edges of the observed regime where the S/N is expected to be lower and in addition, the blueshifted component of \ion{C}{IV} is at most phases beyond the edge of our observed spectrum. Our second method employs fewer features, which, however, have a better S/N and do not suffer from telluric contamination. The derived velocities following both methods are listed in Table \ref{tab04} as in Table \ref{tab03}, whereas the best-fit composite spectra are shown in the two panels of Fig. \ref{fig03}. For brevity, the method that excludes the Br$\gamma$/\ion{He}{I} feature is labeled in the figures and tables as "-Br$\gamma$/\ion{He}{I}", whereas the method that employs only the CNO/\ion{He}{I} blend feature is labeled as "CNO/\ion{He}{I}".

\section{Binary modeling} 
\label{phoebe}

We proceeded to a simultaneous fit of both the $I_{c}$-band light curve and radial velocity curves with a detached, binary model. We used PHOEBE Subversion (release date, 2012$-$07$-$08) \citep{prsa05}, which implements the Wilson-Devinney (WD) code \citep{Wilson71}, to converge into the global optimization using Differential Corrections (DC) powered by a Levenberg-Marquardt fitting scheme. We employed radial velocities measured using FASTWIND templates and the spectra from H05 following both methods discussed in Sec. \ref{hanson}. 

The primary component is defined to be the star eclipsed at phase zero. The period determined from the AoV analysis of the photometric curve was assigned as an initial guess to the period in PHOEBE. We then fit the light curve with the following free parameters: time of primary eclipse HJD$_{0}$, period $P$ (days), inclination $i$ (deg), effective temperature of the secondary component T$_{eff2}$ (K) and surface potentials $\Omega_{1,2}$. We fixed the effective temperature of the primary component at 38 kK. The light curve does not provide evidence for eccentricity, hence we fixed $e=0$. Both components were assumed to be tidally sychronized and surface albedos and gravity brightening exponents were fixed to unity, as for stars with radiative envelopes. Limb darkening coefficients were taken from \cite{Hamme93} using the square-root law. The well-determined period and HJD$_{0}$ were fixed to their converged values and we fit the radial velocity curve for the semi-major axis $\alpha$ (R$_{\sun}$), systemic velocity $\gamma$ (km~s$^{-1}$) and the ratio $q$ of the mass of the secondary component over that of the primary component. In addition, the Rossiter--McLaughlin effect \citep{Ross24} was taken into account to correct for velocity shifts near the conjuction that occur when part of the approaching/receding surface of the occulted star is blocked. The root-mean-square (RMS) value of the fit of each velocity curve was assigned to the "sigma" parameter in PHOEBE, to weight the data in the overall cost function of the analysis. As a next step, both light and radial velocity curves were fit simultaneously, setting free all nine mentioned parameters.

\begin{figure*}
\centering 
\includegraphics[width=6.5in]{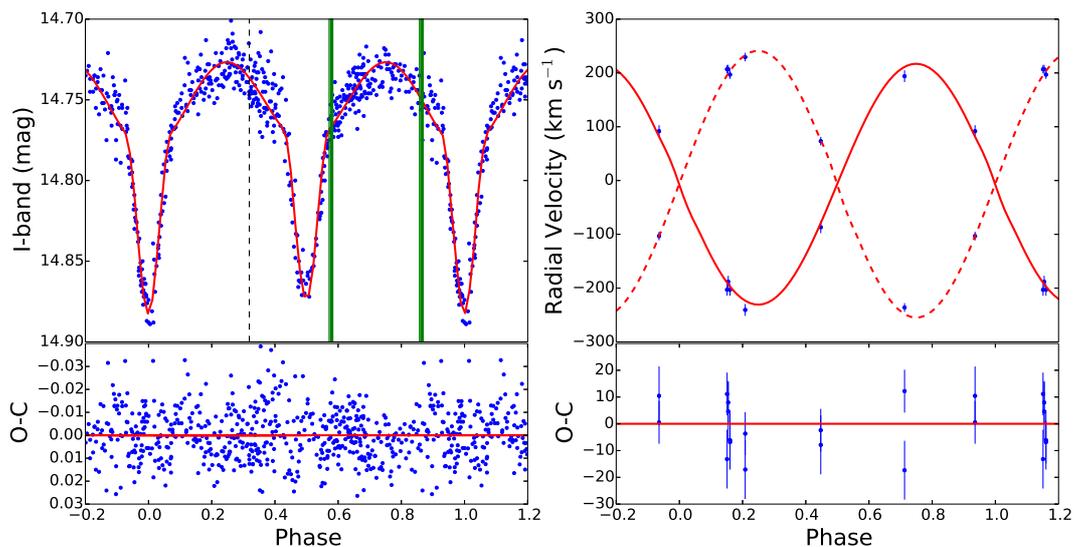}
\caption{The \textit{left panel} shows the $I_{c}$-band light curve of D2-EB, phased to the best-fit parameters derived with PHOEBE using spectra from the H05 atlas. Radial velocities were inferred from the fit of the CNO/\ion{He}{I} blend. The \textit{right panel} shows the radial velocity measurements overplotted by the modeled curve. The solid line denotes the primary and the dashed line the secondary component. Uncertainties correspond to the rms of the best-fit, i.e. 11 km~s$^{-1}$ for the primary and 8 km~s$^{-1}$ for the secondary.  Green, solid lines represent the two observation dates of the $BVI_{c}$ photometry by \cite{Baume09} (six per filter), while the black dashed line indicates the date of the 2MASS measurements. Residuals of the fits are shown in the lower panels.}
\label{fig04}
\end{figure*}

To prevent convergence to a local minimum of the solution space, we used the method of parameter-kicking. We first defined convergence as three consecutive iterations where adjusted parameters are less than or equal to their returned uncertainties. When the criterion is satisfied, all parameters are offset by a relative sigma ("kick"), defined as   \\ 
$$\sigma_\mathrm{kick} = 0.5\dfrac{\left( \chi^2/ N_\mathrm{tot} \right)}{100} $$ \\
\citep{prsa05}, where $N_\mathrm{tot}$ is the total number of points from the fitted photometric and velocity data-sets. Of the 1\,000 sets of parameters, we chose the set that yields the lowest sum of $\chi^2$ values resulting from the fit of both the light and radial velocity curves. We then imported the particular set as input to a new run of 1\,000 iterations, having the parameter-kicking option disabled. The mean and standard deviation of the values for each parameter were derived to provide the solution and the 1-$\sigma$ uncertainties, respectively. The above procedure was repeated three times, for each radial velocity set obtained with FASTWIND and the two methods for the O6.5 III spectrum from H05. The final values are presented in Table \ref{tab05}. It appears that the fit using \ion{He}{II} contributes to a separation of the components that is larger by 1  R$_{\sun}$, compared to the method fitting only on the CNO/\ion{He}{I} feature. In addition, complementing the \ion{He}{II} fit with that of \ion{C}{IV}, yields components of almost equal mass. The systemic velocity modeled with the fit of the synthetic \ion{He}{II} from FASTWIND, is larger by $\sim20$ km~s$^{-1}$ than that modeled with the H05 spectra. The residuals of the three different fits are shown in Tables \ref{tab03} and \ref{tab04}. We found that the velocity measurements based exclusively on the fit of the CNO/\ion{He}{I} blend, yield the best-fit model as can be seen from the resulting RMS values of 11 and 8 km~s$^{-1}$ for the radial velocity curves of the primary and the secondary component, respectively. This best-fit model is displayed in Fig. \ref{fig04}.

The physical parameters of D2-EB are presented in Table \ref{tab06}. All methods resulted in a surface gravity value for both components of log $g \sim 3.9$, which is typical for Galactic O-type dwarfs, whereas effective temperatures of 38 kK and 37 kK indicate spectral types of O6.5 and O7 respectively, with a typical uncertainty of half a spectral type \citep{Martins05}. We adopted a conservative uncertainty of 1kK for the temperatures, equal to the step of the FASTWIND templates used. The inferred light ratio for each method was used to calculate the contribution of the component spectra to the total composite spectrum, as mentioned in Section \ref{rv}. In addition, we derived a negative filling factor $F \sim -1.8$ thus supporting the interpretation that both components are unevolved. The measured radii, period and inclination yield synchronous rotational velocities of $\sim140$ and $\sim130$ km~s$^{-1}$, which are in good agreement with those inferred from the fitting process (Section \ref{fw}). 

Both methods that employ velocities measured from the \ion{He}{II}  line, yield a total mass which is $\sim10$\% greater than the method which discards \ion{He}{II}, owing to the larger separation of the components in the observed spectra. The latter best-fit method provides measurements of the mass and radius of $24.42\pm0.15$ M$_{\sun}$ and $9.69\pm0.07$ R$_{\sun}$ for the primary and $22.06\pm0.68$ M$_{\sun}$ and $8.87\pm0.10$ R$_{\sun}$ for the secondary component. The radii measurements correspond to radii of spheres with equal volume to those defined by the WD surface potentials. Compared to the latest updated catalogue of well-studied, detached EBs by \cite{South14}, the primary component of D2-EB is the third most massive star to be studied with an precision better than $\sim2$\% and the most well-determined star with mass $\ge14.5$ M$_{\sun}$. Nevertheless, we caution that our 0.6\% precision for the mass of the primary is derived as the uncertainty of the solution that optimizes the fit to our observations. Confirming this precision requires more velocity diagnostics and a larger number of observations taken at both quadratures. We hence conclude that the 0.6\% precision should be taken with caution. For this reason, a different code than WD was also applied to evaluate the current solution and revise the uncertainties.

\begin{figure*}
\centering 
\includegraphics[width=6.5in]{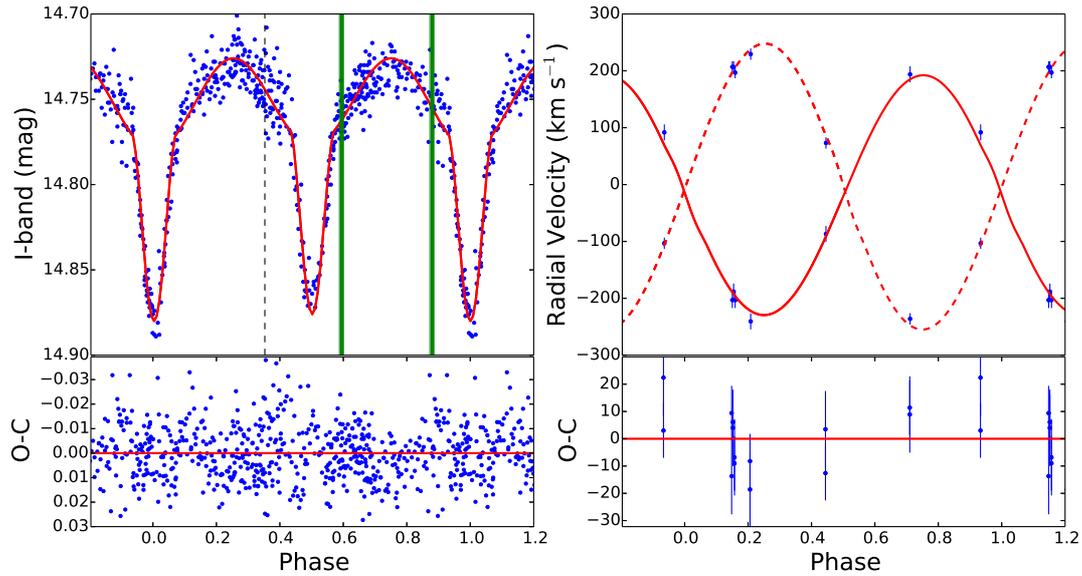}
\caption{Same as Fig. \ref{fig04}, but for the best-fit parameters inferred using ELC. Uncertainties for the radial velocity measurements correspond to 14 km~s$^{-1}$ for the primary and 10 km~s$^{-1}$ for the secondary component.}
\label{fig05}
\end{figure*}

We employed the genetic optimizer of ELC \citep{Oro00} to fit the orbit based on eight input parameters. Specifically, the fit parameters in ELC were HJD$_{0}$, $P$, $i$, the Roche lobe filling factor for each star, $f$, which is the ratio of the radius of the star toward the inner Lagrangian point $L_{1}$ to the distance to $L_{1}$ from the center of the star, $f \equiv x_{\rm point}/x_{L\rm{1}}$, T$_{eff2}$, the primary star's velocity semi-amplitude, $K_{\rm 1}$, and $q$. In fitting the orbit for D2-EB, ELC computed $\sim4\times10^4$ orbits where the values of the eight input parameters varied between fixed ranges judged to be applicable based on the WD fit of the CNO/\ion{He}{I} radial velocity set. The subsequent $\chi^2$ space was then projected as a function of each orbital and astrophysical parameter of interest in the same way as was done in \citet{Wil09}. From the global $\chi^{2}_{min}$, we estimate 1-$\sigma$ uncertainties for derived and fitted parameters from the region where $\chi^2 \leq \chi^{2}_{min} + 1$. These values are listed in Table \ref{tab06}. The inferred masses and radii are $24.5\pm0.9$ M$_{\sun}$ and $9.2\pm0.1$ R$_{\sun}$ for the primary and $21.7\pm0.8$ M$_{\sun}$ and $8.7\pm0.1$ R$_{\sun}$ for the secondary component. The best-fit model with ELC is displayed in Fig. \ref{fig05}.

The masses derived from the best ELC fit are found to be consistent within uncertainties with those derived by PHOEBE and the WD code. The revised uncertainties of the masses of the primary ($3.7\%$) and the secondary ($3.7\%$) are larger than those from our PHOEBE fit while the values of radii are lower by $2-5\%$ than those derived by the WD code and precise to $1\%$. This is not surprising, as the radii rely on the well-constrained light curve, while the masses rely more on the sparse radial velocity data set. We adopt the values from the ELC analysis as their uncertainties are more conservative. 

\section{Distance}
\label{dist}

A prerequisite to precisely determining independent distances to a double-lined eclipsing system is to fit a spectral energy distribution (SED) to accurate, multi-band photometry taken at a known phase. \cite{Baume09} conducted wide field $UBVI_{c}$ observations of the complex that hosts the Danks clusters and their surrounding field and obtained point spread function (PSF) photometry of $\sim$35\,000 sources including our D2-EB\footnote{$B=22.41\pm0.16$ mag, $V=19.40\pm0.02$ mag, $I_{c}=14.73\pm0.03$ mag}. The $U-$band photometry was not available for D2-EB likely due to the high extinction. Using the ephemerides derived by our four methods, we converted the Julian dates of the six per-filter exposures from the \cite{Baume09} data (G.~Carraro, priv. communication) to units of phase. We supplemented the optical photometry with 2MASS measurements \citep{Cutri03} in the \textit{J, H, K$_{s}$}-band obtained on JD 2\,451\,594.871\,3\footnote{$J=11.42\pm0.03$ mag, $H=10.27\pm0.03$ mag, $K_{s}=9.75\pm0.03$ mag}. All observations are consistent within their uncertainties, to out-of-eclipse phases of equal brightness (Figs. \ref{fig04}, \ref{fig05}). To convert the magnitudes to fluxes, we used zeropoints from \cite{Bes98} for the optical photometry and from \cite{Cohen03} for 2MASS.  

SEDs in accordance with the physical parameters of the best-fit spectra were generated with FASTWIND, to provide the flux density per surface unit through our studied bands. The composite flux measured at Earth from a binary at a distance $d$, at a wavelength $\lambda$, reddened to extinction $A(\lambda)$, is given by 

$$ f_{\lambda}=\dfrac{1}{d^2}(R_{1}^{2}F_{1,\lambda}+R_{2}^{2}F_{2,\lambda}) \times 10^{-0.4A(\lambda)} $$ 

\noindent where $R_{1,2}$ and $F_{1,2}$ are the radii and the surface fluxes of the components, respectively. The composite SED was reddened according to the new family of optical and near-infrared extinction laws for O-type stars provided by \cite{Maiz14}, which constitute an improvement of the widely used extinction laws by \cite{Cardelli89}. We ran a fitting algorithm over a wide range of distances with a step of 0.05 kpc, setting free the monochromatic parameters $R_{5495}$ and $E(4405-5495)$ for the type and amount of extinction respectively, and the best-fit values were considered to be those that minimized the weighted-$\chi^2$. To estimate the uncertainty of our measurements, we used a Monte Carlo approach. In particular, we ran the fitting procedure 1\,000 times using sets of randomly selected parameters (photometry and radii) within their uncertainties assuming they are Gaussian-distributed. The corresponding values of distance are shown in Table \ref{tab06} for every set of radii and temperatures determined from the four different fit models. Our adopted radii measured with ELC yielded $d=3.52\pm0.08$ kpc, $E(4405-5495)=3.66\pm0.06$ mag, $R_{5495}=3.26\pm0.04$ and $A_{5495}=11.9\pm0.1$ mag, based on a photometric T$_{eff2}\sim36$ kK. Assuming a spectroscopic T$_{eff2}=37$ kK, the distance changed slightly to $d=3.55\pm0.08$ kpc. Having three degrees of freedom, the reduced-$\chi^2$ of our resulting fits was measured to be $\chi^2_\textrm{red}\sim11$.

\begin{figure*}
\centering 
\includegraphics[width=5in]{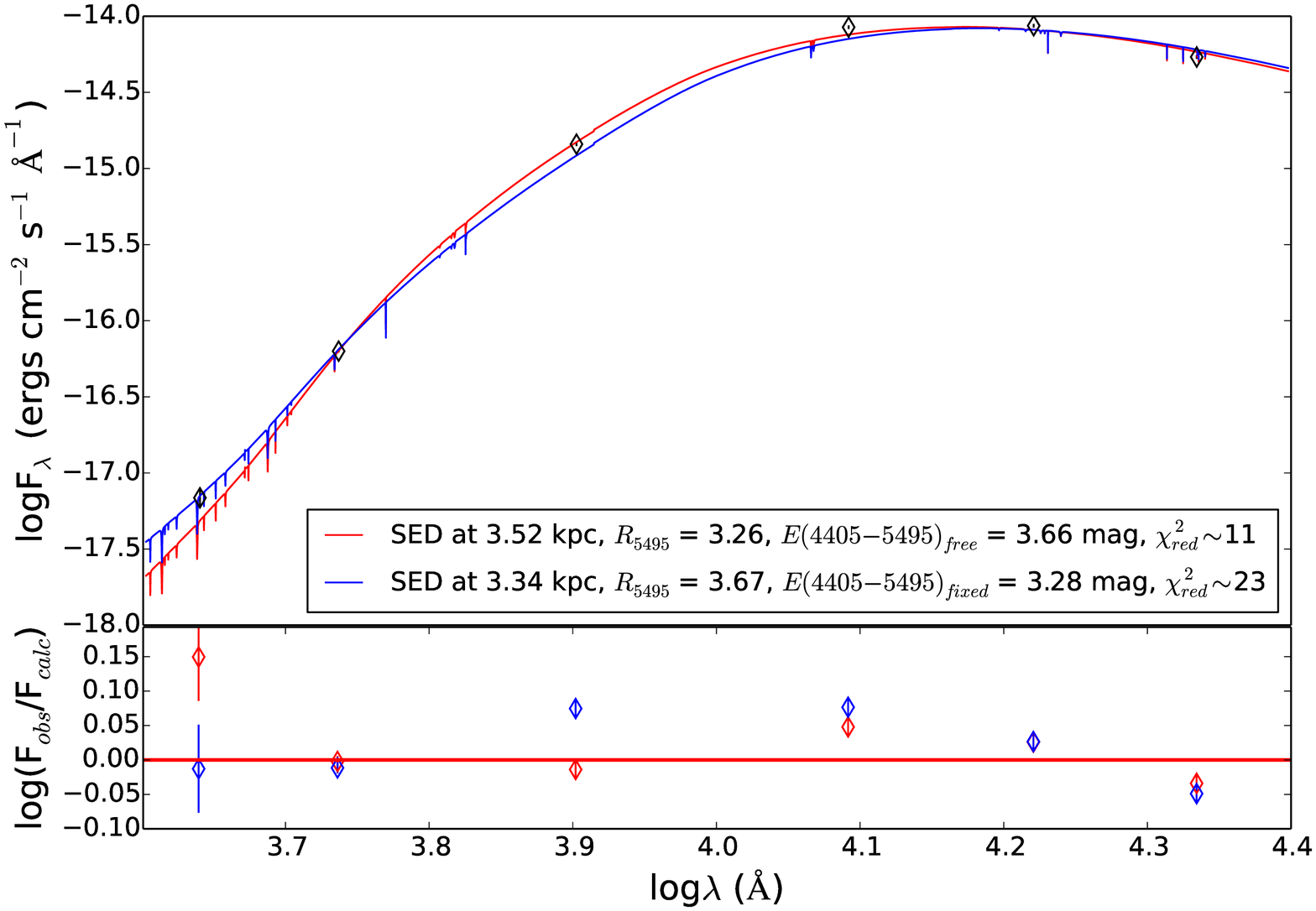}
\caption{Fit of the reddened, composite SEDs using FASTWIND models to the $BVI_{c}JHK_{s}$ photometry, setting the amount of extinction $E(4405-5495)$ as a free parameter (red) and fixed to the value of the observed, band-integrated equivalent $E(B-V)$ (blue). The former method provided the best-fit model yielding $d=3.52\pm0.08$ kpc, $E(4405-5495)=3.66\pm0.06$ mag, $R_{5495}=3.26\pm0.04$ and $A_{5495}=11.9\pm0.1$ mag.}
\label{fig06}
\end{figure*}

We repeated the above procedure setting the amount of extinction $E(4405-5495)$ to the value of its band-integrated equivalent $E(B-V)$ (and so increasing the degrees of freedom by one). Specifically, we calculated $(B-V)_{0}=-0.27$ mag from the synthetic unreddened model at the isophotal wavelengths, which is in agreement with the intrinsic color of O6-9 giants/dwarfs provided by \cite{Martins06}, thus yielding $E(B-V)=3.28\pm0.16$ mag. Our best-fit model then yielded $d=3.34\pm0.12$ kpc, though resulting from a less good fit ($\chi^2_\textrm{red}\sim23$) than that achieved when setting $E(4405-5495)$ free. Both reddened SEDs are shown in Fig. \ref{fig06}. Adopting $E(4405-5495)=3.66$ mag, the residuals indicate a better fit to the \textit{JHK$_{s}$}-band photometry and a reasonable fit to the $I_{c}$-band photometry. The $B$-band photometry clearly deviates from our adopted model causing the discrepancy of $\sim0.4$ mag between the two values of the amount of extinction. This could be partly attributed to the uncertainty of the rather faint $B$-band photometric data ($>22$ mag). However, it may also leave room for a further improvement of the extinction laws. Indeed, having the $B$-band photometry excluded from the fit, our procedure yielded $d=3.53\pm0.08$ and $\chi^2_\textrm{red}\sim14$ (with two degrees of freedom), which still deviates from $\chi^2_\textrm{red}=1$. We caution that in the optical, both families of extinction laws by \cite{Maiz14} and \cite{Cardelli89} have been tested on a sample of low/intermediate reddened stars ($E(4405-5495)<1.5$ mag). Revision of the optical laws for more optically-obscured targets, combined with a different value of the power law exponent for the near-infrared range, could result in a better fit.

\begin{figure*}
\centering 
\includegraphics*[width=4in]{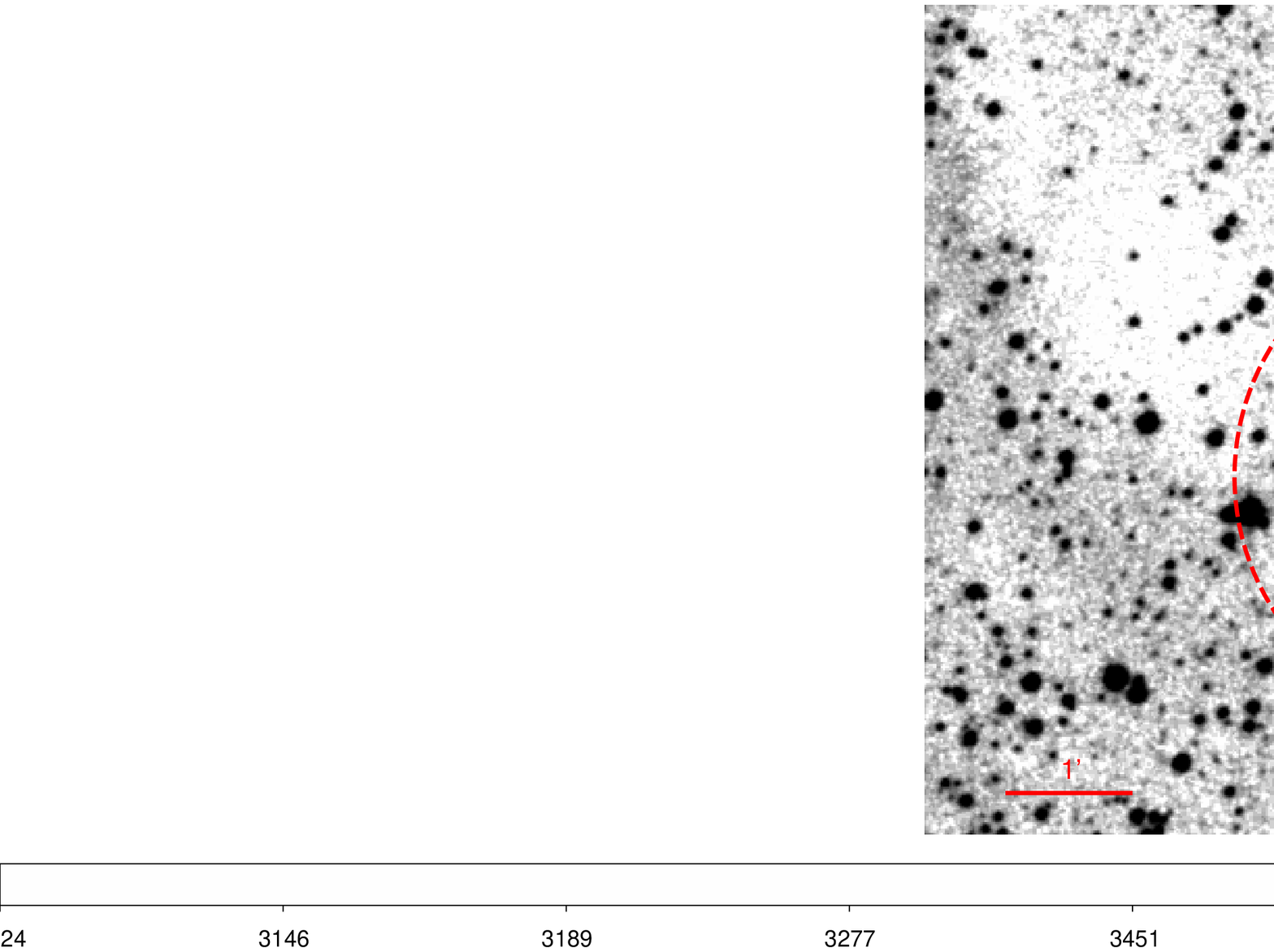}
\caption{Digitized Sky Survey (DSS-2) map of Danks 2 in the red filter. D2-EB (blue arrow) is located outside the $1\farcm5$ radius of the cluster (red solid circle) with an uncertainty of $\pm0\farcm5$ (red dashed circles), as defined by \cite{Chene12}.}
\label{fig07}
\end{figure*}

The measured distance to the D2-EB is in good agreement with reported values of distance to the Danks 2 cluster in the literature: $3.7\pm0.5$ kpc \citep{Chene12}, $3.8\pm0.6$ kpc \citep{Davies11}, $3.4\pm0.2$ kpc \citep{Bica04}. Nevertheless, our inferred, well-constrained physical parameters provide a precision of $\sim2$\%, which is a factor of 3-8 improvement compared to the previous studies. \cite{Baume09} showed that the cluster suffers from substantial reddening with $E(B-V)=2.4$ mag (thus yielding $A_{V}=8.7$ mag), which might be differential across their studied region. D2-EB resides outside their adopted boundary of Danks 2 and likely coincides with molecular gas that contaminates the north-east of the cluster (Fig. \ref{fig07}), which appears to be the reason for our higher value of $A_{5495}=11.9$ mag. We further examined a \textit{Spitzer}/GLIMPSE map at 5.8 $\mu$m and found emission in the region of the gas.

\section{Discussion}
\label{disc}

\begin{figure*}
\centering 
\includegraphics[width=5.5in]{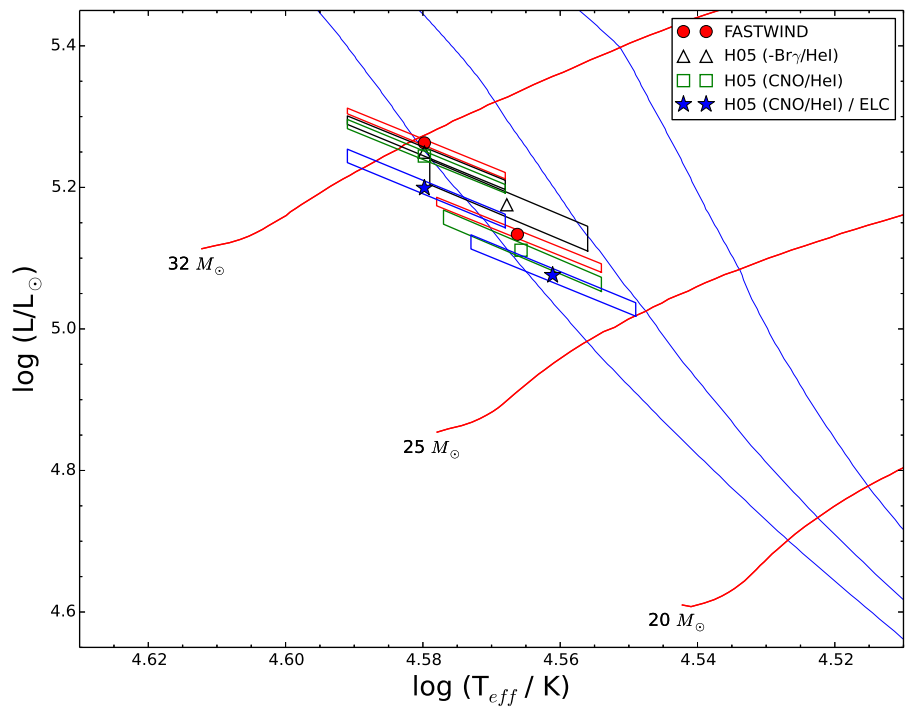}
\caption{H-R diagram showing the comparison of the parameters of D2-EB modeled with the WD code and our three, different sets of radial velocities obtained with: the fit of \ion{He}{II}  of FASTWIND templates (red circles), the fit of all $K-$band features of H05 spectra except for the Br$\gamma$/\ion{He}{I} blend (black triangles), the fit of the CNO/\ion{He}{I} blend of the H05 spectra (green squares). The last set is also modeled with ELC and parameters are shown with blue asterisks.
Evolutionary tracks and isochrones for single stars \citep{Ekstrom12} at $Z=0.014$ with rotation, are shown. Isochrones from left to right, correspond to 3.2, 4 and 5 Myr. All methods yield co-evolutionary components with an age of $\sim3.2$ Myr that appear overluminous for their masses.}
\label{fig08}
\end{figure*}

To measure the age of D2-EB and therefore the Danks 2 cluster, we used the evolutionary models of \cite{Ekstrom12} for single stars, at $Z=0.014$. The models assume initial stellar rotation at 40\% of the critical velocity accounting for effects discussed in \cite{Meynet00}, and correspond to main sequence velocities of $110-220$ km~s$^{-1}$, which are suitable for our case. All methods followed to derive masses agree with an age of $\sim3.2$ Myr for both components, according to the Hertzsprung–Russell (H-R) diagram shown in Fig. \ref{fig08}. A by-eye interpolation of the evolutionary tracks implies that the members of D2-EB appear overluminous for their masses. In particular, using \ion{He}{II}  for measuring velocities, the obtained masses found are $\sim15$\% lower than those predicted, for both components. When excluding \ion{He}{II} from the velocity analysis, this mass discrepancy increases to $\sim20$\%. A mass discrepancy is often reported when dealing with contact and semidetached massive binary systems \citep{Burk97,Bonanos09,Koumpia12} where one or both components are ovefilling their Roche lobe and is clearly attributed to mass transfer. Nevertheless, in the case of the well-detached D2-EB, we have no evidence for Roche lobe overflow and in addition, both components appear less massive than expected. Similar cases of mass discrepancy in detached systems have been also reported by \cite{Gonz05}, \cite{Wil08} and most recently by \cite{Massey12}. The latter study concluded that models with enhanced convective overshooting or higher initial rotation should be adopted in order to explain the $\sim$11\% mass discrepancy observed in two detached systems in the Large Magellanic Cloud. A recent study by \cite{Markova15} showed that a mass discrepancy for Galactic O stars with initial mass $<$ 35 M$_{\sun}$ is evident, with the spectroscopically derived masses being systematically lower than the evolutionary masses inferred from the theoretical models. Stars with spectroscopically measured masses of $\sim25$ M$_{\sun}$ and $\sim22$ M$_{\sun}$, equal to our adopted values for the components of D2-EB, were shown to display evolutionary masses of $\sim30$ M$_{\sun}$ and $\sim27$ M$_{\sun}$ respectively, which are in a good agreement with our predictions from the theoretical models.    

\begin{figure*}
\centering 
\includegraphics[width=5.5in]{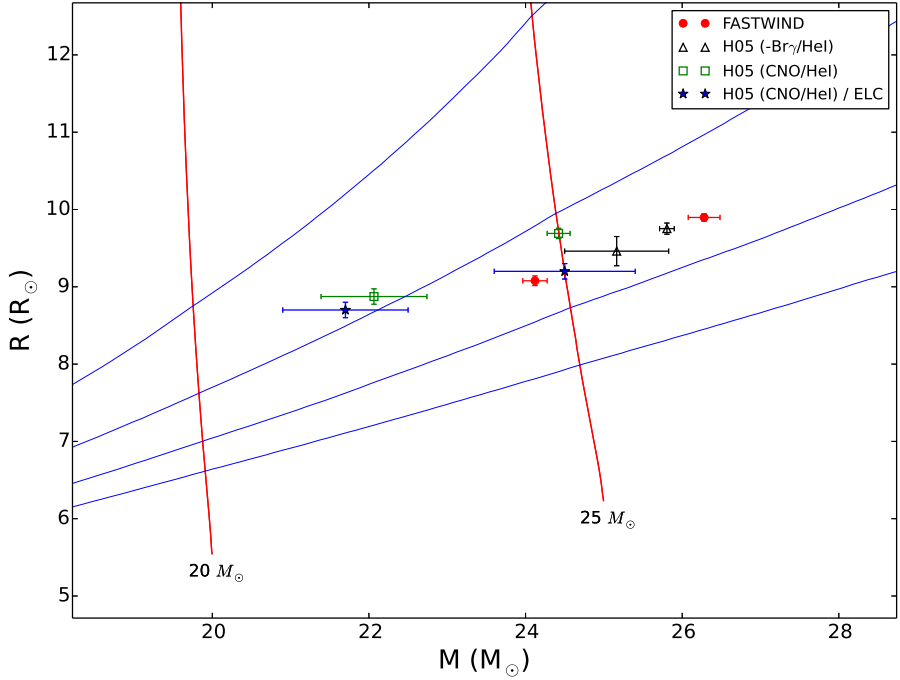}
\caption{Mass-Radius (M-R) diagram showing the comparison of the parameters of D2-EB with evolutionary tracks and isochrones for single stars \citep{Ekstrom12} at $Z=0.014$ with rotation. The symbols denoting each method are the same as in Fig. \ref{fig08}. Isochrones from the bottom up, correspond to 3.2, 4, 5 and 6.3 Myr. We find that both methods based on a \ion{He}{II}  fit for measuring velocities yield co-evolutionary components of age 4.5 Myr. The best-fit model based on the CNO/\ion{He}{I} blend fit yields an age of $\sim5$ Myr, as modeled both with the WD code and ELC.}
\label{fig09}
\end{figure*}

\cite{Davies11} suggested an age for the cluster of $2-6$ Myr, based on the presence of the carbon-rich Wolf-Rayet D2-3 and the luminosity of the brightest OB supergiant in Danks 2. \cite{Chene12} fit theoretical isochrones to a colour-magnitude diagram of their near-infrared photometry and argued that Danks 2 is likely older, with an age of $4-7$ Myr. A temperature-independent method to evaluate the evolutionary status of D2-EB is the use of the mass-radius diagram, presented in Fig. \ref{fig09}. The two methods based on the fit of \ion{He}{II} are both consistent with an age of 4.5 Myr. Our adopted, best-fit model yields an age of $\sim5$ Myr, following from both the WD code or ELC. All measured values of age are found to be within the range of ages determined for Danks 2 by the previous studies. 

D2-EB resides slightly outside the $1\farcm5\pm0\farcm5$ angular size of Danks 2 \citep[][see Fig. \ref{fig07}; solid and dashed circles]{Chene12} beyond which, the cluster density profile falls below that of the field. The systemic velocity ($-7\pm1$ km~s$^{-1}$) is not in agreement with the radial velocity of Danks 2 \citep[$-44\pm8$ km~s$^{-1}$,][]{Chene12} nor the host G305 complex \citep[$-39\pm3$ km~s$^{-1}$,][]{Davies11}. This suggests that D2-EB may have been ejected from the cluster as a runaway binary. The radial velocity of D2-EB relative to the cluster is $37\pm8$ km~s$^{-1}$. This can be taken as a lower limit for the space velocity at which the system is escaping from the cluster. The angular distance 2\farcm1 from the center corresponds to a projected linear distance of $2.2\pm0.1$ pc. Assuming a kinematic age $\sim5$ Myr equal to the measured age of the system, the tangential velocity would be less than 1 km~s$^{-1}$.

\cite{Blaaw61} hypothesized that runaway stars are created when a massive component within a binary system explodes as a Type II supernova. The secondary is then ejected with a velocity comparable to the orbital velocity at the time of the supernova event. The explosion may not disrupt the system \citep{Hills83}, which will be observed as a OB-neutron star/black-hole system and eventually become a high-mass X-ray binary. Given that D2-EB is a double-lined O-type binary, it is unlikely that such a mechanism took place. Alternatively, D2-EB could have been part of a triple system including a very massive component ($\ge85$ M$_{\sun}$) with a lifetime of less than $4.5$ Myr. However, the inferred radial velocity curve lacks evidence of a low-mass third companion, unless it orbits with a high eccentricity and/or long period. Observations of D2-EB over a longer time span than the data presented here are required to investigate the possibility of a third body in the system. \cite{Poveda67} suggested that runaway stars are dynamically ejected due to encounters of collapsing protostars in/near the core of young clusters. Interactions in clusters that contain initial "hard" binaries increase the number of escapees via binary-binary interactions \citep{Mikkola83}. The less massive binary system reaches peculiar space velocities up to $\sim200$ km~s$^{-1}$, while the more massive system travels at a speed less than $\sim100$ km~s$^{-1}$ \citep{Leonard88}. The binary frequency for runaways with V$_{\infty}>30$ km~s$^{-1}$ is predicted to be 10\% and it is striking that systems hosting components of $M\sim20$  M$_{\sun}$ as the most massive members are not predicted to escape with more than 50 km~s$^{-1}$ \citep{Leonard90}. This does not conflict with the observed lower limit for the space velocity of D2-EB. A young, dense cluster core increases the possibility of dynamical ejection, with the consequence that the kinematic age of the runaways is similar to the age of the cluster \citep{Gual04}. In this case, the unreasonably low estimated value for the tangential velocity of D2-EB renders the possibility of the dynamical ejection through binary-binary interaction to be unlikely.

\section{Conclusions}
\label{concl}

We present an analysis of new \textit{K}-band spectra from VLT/ISAAC with which we determined accurate fundamental parameters of D2-EB, a massive, early-type eclipsing binary in the young cluster Danks 2, which is embedded in the G305 Galactic, star-forming region. The best-fit model to the binary was obtained by using the \ion{He}{I} line ($2.112$ $\mu$m) and the CNO complex for measuring radial velocities. The system was found to contain two co-evolutionary O6.5-7 main-sequence components with an age of $\sim5$ Myr. We determined masses and radii of $24.5\pm0.9$ M$_{\sun}$ and $9.2\pm0.1$ R$_{\sun}$ for the primary and $21.7\pm0.8$ M$_{\sun}$ and $8.7\pm0.1$ R$_{\sun}$ for the secondary component with a precision $\sim3.8$\% for the masses and $\sim1$\% for the radii. Models utilising a fit of \ion{He}{II} for measuring velocities, were found to yield a $\sim10$\% higher total mass and $\sim0.5$ Myr younger age.

Employing the precise measurements of the radii and out-of-eclipse optical and near-infrared photometry of D2-EB, we determined a distance to the system of $d=3.52\pm0.08$ kpc from a fit to the SED of the system. Up to now, this is the most well-constrained distance measurement to the Danks clusters and thus to the host complex, with a precision of $\sim2$\%. D2-EB is found to reside slightly outside the cluster and is optically obscured by molecular gas, which causes an extinction for the system of $A_{5495}=11.9\pm0.1$ mag. We report a systemic velocity for the system that is inconsistent with that of Danks 2, making D2-EB a candidate runaway binary. However, neither of the two main mechanisms thought to yield runaways provide a robust explanation of the nature of the ejection of D2-EB from the Danks 2 cluster. Future proper motion measurements with Gaia\footnote{http://www.cosmos.esa.int/web/gaia/science-performance}, which will be complete down to 20th mag in the visual, should constrain the motion of both the cluster and the system, and provide well-determined values of the space velocity and the kinematic age of D2-EB.

\begin{acknowledgements}
MK and AZB acknowledge funding by the European Union (European Social Fund), National Resources under the "ARISTEIA" action of the Operational Programme "Education and Lifelong Learning" in Greece. MK, AZB and EK acknowledge research and travel support from the European Commission Framework Program Seven under the Marie Curie International Reintegration Grant PIRG04-GA-2008-239335. Support for JLP is in part by FONDECYT through the grant 1151445 and by the Ministry of Economy, Development, and Tourisms Millennium Science Initiative through grant IC120009, awarded to The Millennium Institute of Astrophysics, MAS. This publication makes use of data products from the Two Micron All Sky Survey, which is a joint project of the University of Massachusetts and the Infrared Processing and Analysis Center/California Institute of Technology, funded by the National Aeronautics and Space Administration and the National Science Foundation. The Digitized Sky Surveys were produced at the Space Telescope Science Institute under U.S. Government grant NAG W-2166. The images of these surveys are based on photographic data obtained using the Oschin Schmidt Telescope on Palomar Mountain and the UK Schmidt Telescope. The plates were processed into the present compressed digital form with the permission of these institutions. This research has made use of the VizieR catalogue
access tool, CDS, Strasbourg, France. 
\end{acknowledgements}

\clearpage

\begin{table*}

\caption[]{Summary of the parameters of the Danks 2 cluster and D2-EB.}
\label{tab01}

\begin{tabular}{l|cc}

\hline \hline
\quad\quad Parameter & Danks 2  &  D2-EB (this work)   \\ 
\hline

Total Mass (M$_{\sun}$)  &  $3\,000\pm800$$^{1}$  &    $46.2\pm1.2$       \\                
Age (Myr)                &        $4-7$$^{2}$      &       $\sim5$         \\                
Distance (kpc)           &     $3.4\pm0.2$$^{3}$   &       $3.52\pm0.08$   \\                
Visual Extinction (mag)  &        8.7$^{4}$        &         11.9          \\                

\hline
\end{tabular}

\textbf{References:} $^{1}$\cite{Davies11}, $^{2}$\cite{Chene12},\\ $^{3}$\cite{Bica04}, $^{4}$\cite{Baume09}

\end{table*}

\begin{table*}

\caption[]{Log of spectroscopic observations of D2-EB with VLT/ISAAC.}
\label{tab02}  

{\centering
\begin{tabular}{l|cc}

\hline \hline
\quad\quad HJD & Total integration time (s)  &    Airmass  \\ 
\hline

2\,456\,340.782\,01  &  2x100       &       1.31   \\                
2\,456\,345.856\,71  &  2x100       &       1.28   \\                
2\,456\,360.828\,68  &  2x100       &       1.30   \\                
2\,456\,360.840\,70  &  2x100       &       1.31   \\                
2\,456\,360.859\,82  &  2x100       &       1.35   \\                
2\,456\,378.695\,33  &  2x100       &       1.29   \\                
2\,456\,378.808\,74* &  2x100       &       1.34  \\                 
2\,456\,383.715\,90  &  2x100       &       1.27   \\                
\hline
\end{tabular} 
} 

(*) Discarded from our analysis.

\end{table*} 

\begin{table*}
{\tiny
\caption[]{Radial velocity measurements of D2-EB using FASTWIND models.}
\label{tab03}  

{\centering
\begin{tabular}{c|crrrr}

\hline \hline
   & \multicolumn{5}{c}{FASTWIND}  \\

HJD      & Orbital & RV$_{1}\quad $ & (O$-$C)$_{1}$ & RV$_{2}\quad $ & (O$-$C)$_{2}$ \\
         & phase & (km~s$^{-1}$)  & (km~s$^{-1}$)  & (km~s$^{-1}$)  & (km~s$^{-1}$) \\

\hline
2\,456\,340.782\,01  & 0.19  & $-$$181\pm19$  &    24 &    $279\pm30$ &    32    \\
2\,456\,345.856\,71  & 0.69  &    $229\pm19$  & $-$1  & $-$$171\pm30$ &    55    \\
2\,456\,360.828\,68  & 0.13  & $-$$173\pm19$  & $-$12 &    $207\pm30$ &     8    \\
2\,456\,360.840\,70  & 0.14  & $-$$183\pm19$  & $-$19 &    $182\pm30$ & $-$21    \\
2\,456\,360.859\,82  & 0.14  & $-$$193\pm19$  & $-$23 &    $217\pm30$ &     8    \\
2\,456\,378.695\,33  & 0.43  & $-$$117\pm19$  & $-$27 &     $88\pm30$ & $-$34    \\
2\,456\,383.715\,90  & 0.92  &    $142\pm19$  &    13 &  $-$$98\pm30$ &    18    \\
\hline
\end{tabular} 
} 

} 
\end{table*}

\begin{table*}
{\tiny
\caption[]{Radial velocity measurements of D2-EB using spectra from H05.}
\label{tab04}  

{\centering
\begin{tabular}{c|crrrr|crrrr}

\hline \hline
   & \multicolumn{5}{c|}{H05 (-Br$\gamma$/\ion{He}{I})}  & \multicolumn{5}{c}{H05 (CNO/\ion{He}{I})} \\

HJD      & Orbital & RV$_{1}\quad $ & (O$-$C)$_{1}$ & RV$_{2}\quad $ & (O$-$C)$_{2}$ & Orbital & RV$_{1}\quad $ & (O$-$C)$_{1}$ & RV$_{2}\quad $ & (O$-$C)$_{2}$ \\
         & phase & (km~s$^{-1}$)  & (km~s$^{-1}$)  & (km~s$^{-1}$)  & (km~s$^{-1}$) & phase & (km~s$^{-1}$)  & (km~s$^{-1}$)  & (km~s$^{-1}$)  & (km~s$^{-1}$)\\
\hline
2\,456\,340.782\,01  & 0.20  &  $-$$241\pm14$   &      1   &    $209\pm11$   &  $-$12  & 0.21 & $-$$241\pm11$ & $-$17 &    $229\pm8$ &  $-$4 \\
2\,456\,345.856\,71  & 0.70  &     $189\pm14$   &  $-$29   & $-$$241\pm11$   &      8  & 0.71 &    $194\pm11$ & $-$17 & $-$$236\pm8$ &    12 \\
2\,456\,360.828\,68  & 0.14  &  $-$$213\pm14$   &  $-$11   &    $197\pm11$   &     17  & 0.15 & $-$$203\pm11$ & $-$13 &    $207\pm8$ &    11 \\
2\,456\,360.840\,70  & 0.15  &  $-$$223\pm14$   &  $-$18   &    $197\pm11$   &     13  & 0.15 & $-$$188\pm11$ &     5 &    $207\pm8$ &     8 \\
2\,456\,360.859\,82  & 0.15  &  $-$$213\pm14$   &  $-$3    &    $182\pm11$   &   $-$7  & 0.16 & $-$$203\pm11$ &  $-$6 &    $197\pm8$ &  $-$7 \\
2\,456\,378.695\,33  & 0.44  &  $-$$117\pm14$   &  $-$14   &     $83\pm11$   &      3  & 0.45 &  $-$$87\pm11$ &  $-$8 &     $73\pm8$ &  $-$2 \\
2\,456\,383.715\,90  & 0.93  &     $107\pm14$   &     14   & $-$$113\pm11$   &      8  & 0.94 &     $92\pm11$ &    10 & $-$$103\pm8$ &     1 \\
\hline
\end{tabular} 
} 

} 
\end{table*}

\begin{table*}
{\tiny

\caption[]{Results from the analysis of the light and radial velocity curves with PHOEBE.}
\label{tab05} 

\begin{tabular}{lccc}

\hline \hline

Parameter & FASTWIND & H05 (-Br$\gamma$/\ion{He}{I}) & H05 (CNO/\ion{He}{I})\\
\hline
Period, $P$ (days)                 &      $3.373\,652\pm0.000\,002$       &      $3.373\,474\pm0.000\,003$      &         $3.373\,335\pm0.000\,003$      \\
Time of primary eclipse, HJD$_{0}$ &  $2\,455\,685.653\,33\pm0.000\,1$         & $2\,455\,685.652\,73\pm0.000\,1$        &    $2\,455\,685.652\,45\pm0.000\,1$        \\
Inclination, $i$ (deg)             &          $68.4\pm0.2$            &          $68.0\pm0.2$           &             $68.2\pm0.2$           \\
Surface potential, $\Omega_{1}$    &          $4.50\pm0.02$           &          $4.63\pm0.05$          &             $4.47\pm0.03$          \\
Surface potential, $\Omega_{2}$    &          $4.62\pm0.04$           &          $4.68\pm0.05$          &             $4.56\pm0.08$          \\
Light ratio in I, $L_{2}/L_{1}$    &          $0.79\pm0.01$           &          $0.89\pm0.05$          &             $0.79\pm0.01$          \\
Mass ratio, $q$                    &          $0.92\pm0.01$           &          $0.98\pm0.02$          &             $0.90\pm0.02$          \\ 
Systemic velocity, $\gamma$ (km~s$^{-1}$) &     $11\pm1$              &         $-$$13\pm1$             &             $-$$7\pm1$             \\
Semi-major axis, $\alpha$ (R$_{\sun}$)    &  $34.95\pm0.04$           &         $35.08\pm0.17$          &            $34.02\pm0.20$          \\
Semi-amplitude, $K_{1}$ (km~s$^{-1}$) &        $233\pm1$              &          $241\pm4$              &              $225\pm5$             \\
Semi-amplitude, $K_{2}$ (km~s$^{-1}$) &        $254\pm1$              &          $247\pm2$              &              $249\pm3$             \\
Radius, r$_{1,pole}$               &         $0.276\pm0.002$          &         $0.271\pm0.002$         &            $0.278\pm0.002$         \\
............ r$_{1,side}$          &         $0.282\pm0.002$          &         $0.277\pm0.002$         &            $0.284\pm0.002$         \\
............ r$_{1,point}$         &         $0.295\pm0.003$          &         $0.290\pm0.003$		&            $0.297\pm0.003$         \\
............ r$_{1,back}$          &         $0.290\pm0.003$          &         $0.285\pm0.002$         &            $0.292\pm0.002$         \\
Radius, r$_{2,pole}$               &         $0.254\pm0.002$          &         $0.264\pm0.002$         &            $0.255\pm0.002$         \\
............ r$_{2,side}$          &         $0.259\pm0.003$          &         $0.269\pm0.002$         &            $0.260\pm0.002$         \\
............ r$_{2,point}$         &         $0.269\pm0.003$          &         $0.280\pm0.003$         &            $0.271\pm0.003$         \\ 
............ r$_{2,back}$          &         $0.266\pm0.003$          &         $0.276\pm0.003$         &            $0.267\pm0.003$         \\
\hline

\end{tabular} 
} 
\end{table*}

\begin{table*}
\caption[]{Physical parameters of D2-EB.}
\label{tab06} 
\begin{tabular}{lcccc}

\hline \hline
Parameter & FASTWIND & H05 (-Br$\gamma$/\ion{He}{I}) & H05 (CNO/\ion{He}{I})  & H05 (CNO/\ion{He}{I}) (ELC) \\
\hline

$M_{1}$ ( M$_{\sun}$)  & $26.28\pm0.20$    &  $25.80\pm0.09$  &  $24.42\pm0.15$  &    $24.5\pm0.9$  \\
$M_{2}$ ( M$_{\sun}$)  & $24.12\pm0.16$    &  $25.16\pm0.66$  &  $22.06\pm0.68$  &    $21.7\pm0.8$  \\
$R_{1}$ ( R$_{\sun}$)  &  $9.90\pm0.05$    &  $9.75\pm0.07$   &   $9.69\pm0.07$  &     $9.2\pm0.1$  \\
$R_{2}$ ( R$_{\sun}$)  &  $9.08\pm0.06$    &  $9.46\pm0.19$   &   $8.87\pm0.10$  &     $8.7\pm0.1$  \\
log $g_{1}$           & $3.87\pm0.01$     &  $3.87\pm0.01$   &   $3.85\pm0.01$  &    $3.90\pm0.01$ \\
log $g_{2}$           & $3.90\pm0.01$     &  $3.89\pm0.01$   &   $3.89\pm0.01$  &    $3.90\pm0.01$ \\
$T_{eff1}$ (K)     &                      \multicolumn{4}{c}{38000 (fixed)}                     \\
T$_{eff2}$ (K)     &$36833\pm1000$        & $36964\pm1000$   &  $36788\pm1000$  &   $36400\pm1000$ \\
log $L_{1}/L_{\sun}$  & $5.26\pm0.05$     &  $5.25\pm0.05$   &   $5.24\pm0.05$  &    $5.20\pm0.05$ \\
log $L_{2}/L_{\sun}$  & $5.13\pm0.05$     &  $5.18\pm0.05$   &   $5.11\pm0.05$  &    $5.08\pm0.05$ \\
Distance (kpc)        & $3.77\pm0.08$     &  $3.81\pm0.09$   &   $3.68\pm0.08$  &    $3.52\pm0.08$ \\
\hline

\end{tabular} 

\end{table*}

\end{document}